# Explaining atomic clock behavior in a gravitational field with only 1905 Relativity

**Rafael A. Valls Hidalgo-Gato** [1]
**Natalio Svarch Scharager** [2]

**Abstract:** Supported only in the two 1905 Einstein's papers on Relativity and a very rigid respect for the historical context, an analysis is done of the derivation of the universal mass-energy relationship. It is found, contrary to the today accepted Physics knowledge, that a body's **Rest Mass** measures its **Potential Energy** in the 1905 context. After emphasizing the difference between **1905 Relativity** (**1905R**) and **Special Relativity** (**SR**), the developing of a **1905R** relativistic gravity is started for a small mass **m** material point moving in the central gravitational field of a great mass **M** one. A formula for the rest mass $\mathbf{m_0}$ as a function of its distance **r** from **M** is obtained. Finally, those results are applied to an atomic clock in a gravitational field**,** reaching a factor to obtain the clock time rate change very close to the **GR** one. The factors from **1905R** and **GR** are compared, emphasizing the absent of a singularity in **1905R**. In the conclusions, a new road for the development of a 1905R relativistic mechanics is declared, related with the discovery that **Rest Mass** measures **Potential Energy**, done by 1905 Einstein even if not realizing it.

## 1. Introduction.

The change in time rate of an atomic clock owed to a change in speed or position in a gravitational field are well-known phenomena in today Physics. The change owed to speed was already correctly predicted by **1905 Relativity** since A. Einstein's first paper [1] on the topic, but the one owed to a change of position in a gravitational field was explained only about a decade later applying his new developed **General Relativity** (**GR**) theory [2].

In paper [2], all relativity before **GR** is denoted **Special Relativity** (**SR**) to distinguish it from **GR** (hereafter in this article, between " " denotes 1905 Einstein's literal words; "The theory which is presented in the following pages conceivably constitutes the farthest-reaching generalization of a theory which, today, is generally called the 'theory of relativity'; I will call the latter one –in order to distinguish it from the first named- the 'special theory of relativity', which I assume to be known.", at the beginning of paper [2]), being considered since then that **gravity** is excluded from **SR**. Interesting enough, the example that appears at the end of **§ 4** in paper [1] refers to our real rotating Earth with a gravitational centripetal accelerated clock at its equator and another at rest in one pole. Here is precisely where the prediction mentioned before is done, running the moving clock at the equator a little slow that the rest one "at one of the poles under otherwise identical conditions".

[1] Institute of Cybernetics, Mathematics and Physics (ICIMAF), CITMA, Cuba.
email: valls@icmf.inf.cu
[2] Havana's University Superior Institute of Medical Sciences, MINSAP, Cuba.

Fortunately, the today Earth Centered Inertial (ECI) system of the Global Positioning System (GPS) [3] corresponds exactly with the one being denoted "stationary" by 1905 Einstein ("a system of co-ordinates in which the equations of Newtonian mechanics hold good", at the beginning of **§ 1** in paper [1]). The "moving" system is the clock at the rotating equator.

As we can see, gravity is present in a protagonist role in the1905 Einstein's example. The facts since the Hafele&Keating [4, 5] experiment, and overall, the huge experimental evidence of the very successful GPS [3], put out of any doubt that the presence of gravity is not an impediment for the holding of 1905 Relativity formulas. An interesting question arises in a natural way here:

Was the development of **General Relativity** the unique valid alternative for the explanation of atomic clock behavior in a gravitational field?

The goal of this paper, already advanced in its title, is to explain atomic clock behavior using only **1905 Relativity** (hereafter denoted **1905R**), considering **1905R** only the papers published in that year. It is then very important here to distinguish **1905R** from **SR**. All the development of relativity after 1905 up to the development of **GR** is considered out from **1905R**, among them the united space-time concept and the use of 4-tensors. The 1907 H. Minkowski's work in relativity is considered out from **1905R**.

## 2. Analyzing the 1905R derivation of the mass-energy relationship.

Only about three months after his first **1905R** paper, Einstein writes his second one [6], very short indeed, with only 3 pages. Here he derives the mass-energy relationship in a so general way that we can denote it as a really universal one.

When analyzing any old text, we must be careful for not introducing any concept developed in the future of its writing epoch. That implies a not valid interpretation, because no thing can be used before its creation moment. In the 1905 September 27 text (paper [6]), the deduction of the mass-energy relationship is started denoting $\mathbf{E_0}$ the rest energy of a body considered stationary in some inertial system $\mathbf{S_1}$. In today Physics any **Rest Energy** is related with a constant intrinsic body's **Rest Mass**, not depending on its position or velocity (and then having no relation at all with potential or kinetic energy). That concept is developed evidently after the discovery of the mass-energy relationship, being then totally invalid to use it when interpreting precisely the old paper where that relationship is derived.

After denoting $\mathbf{H_0}$ the energy of the same body moving with velocity $\mathbf{v}$ in another inertial system $\mathbf{S_2}$, the derivation continues with the emission by the body of some quantity $\mathbf{L}$ of light energy divided in two equal parts $\mathbf{L/2}$, as viewed in $\mathbf{S_1}$, in such a way that the body remains at rest in that system. The state before the emission is denoted by the sub-index **0**, and after the emission by the sub-index **1**.

Using a formula derived in paper [2] that relates the energies of some quantity of light as viewed in two different inertial systems, the Conservation of Energy Principle and his 1905 Principle of Relativity ("the same laws of electrodynamics and optics will be valid for all frames of reference for which the equations of mechanics hold good", at the Introduction of paper [1]), 1905 Einstein reaches the following equations

$$\mathbf{E_0 = E_1 + L} \qquad \mathbf{(2.1)}$$
$$\mathbf{H_0 = H_1 + L\ [1/\sqrt{(1 - v^2/c^2)}]\ ,} \qquad \mathbf{(2.2)}$$

and subtracting them to the following one

$$\mathbf{H_0 - E_0 - (H_1 - E_1) = L\ \{[1/\sqrt{(1 - v^2/c^2)}] - 1\}} \ \mathbf{(2.3)}$$

Let us consider the following crucial text of the 27Sep1905 Einstein's paper:

"The two differences of the form **H – E** occurring in this expression have simple physical significations. **H** and **E** are energy values of the same body referred to two systems of co-ordinates ($\mathbf{S_2}$ and $\mathbf{S_1}$ respectively) which are in motion relatively to each other, the body being at rest in one of the two systems ($\mathbf{S_1}$). Thus it is clear that the difference **H – E** can differ from the kinetic energy **K** of the body, with respect to the other system $\mathbf{S_2}$, only by an additive constant **C**, which depends on the choice of the arbitrary additive constants of the energies **H** and **E**. Thus we may place

$$\mathbf{H_0 - E_0 = K_0 + C,} \qquad \mathbf{(2.4)}$$
$$\mathbf{H_1 - E_1 = K_1 + C,} \qquad \mathbf{(2.5)}$$

since **C** does not change during the emission of light. So we have

$$\mathbf{K_0 - K_1 = L\ \{[1/\sqrt{(1 - v^2/c^2)}] - 1\}\text{"}} \qquad \mathbf{(2.6)}$$

Interpreting this text in its 1905 historical context, putting the expressions **H – E = K + C** as **H = K + (E + C)**, we can see easily in them the Conservation Principle of Energy

**Total Energy (H) = Kinetic Energy (K) + Potential Energy (E+C) (2.7)**

with the presence of the arbitrary additive constant **C** characteristic of **Potential Energy**.

We haven't in the 1905 context any other kind of energy **E** different from the potential one with an arbitrary additive constant **C**. We also haven't any other expression for the Conservation Principle of Energy different from **(2.7)**. It is then completely clear in the 1905 context that a body rest energy **E** can be only its **Potential Energy**. It can never be interpreted as a **Rest Energy** not depending on position or velocity, by a very simple and trivial logical reason: that concept doesn't exist yet in the writing moment. The classical definition of **Potential Energy** is precisely the energy a body has owed to its position, being its **Total Energy** if the body is at rest (kinetic energy equal zero). Without any doubt at all, in the paper [1] **E** can be only the body's **Potential Energy**.

## 3. Rest Mass measures Potential Energy in 1905R.

The principal conclusion of the 1905 September 27 Einstein's paper is that "The mass of a body is a measure of its energy-content". From that general conclusion we can derive the following particular one (when the body is at rest):

"The (**rest**) mass of a body is a measure of its (**potential**) energy-content".

The confirmation of the presence of **Potential Energy** in **1905R** is really a surprising fact. More than a century had elapsed since that, full of Physics achievements. Among them, the development of **General Relativity (GR)** as the more advanced gravity theory (the basis of modern Cosmology), and **Quantum Mechanics** (**QM**) totally consolidated explaining molecules, atoms and elemental particles.

However, it is known that conflicts between **GR** and **QM** persist until today, even being Einstein a pioneer in the development of both theories. Maybe the knowledge of **Rest Mass** measuring **Potential Energy** can helps in some way in this complex topic? For the moment, we will limit ourselves to the **1905R** context. This implies to take special care in not to use concepts belonging to **SR**, **GR, QM,** or any other theory developed after **1905R**.

## 4. Developing a 1905R relativistic gravity.

To fix the historic context in which the following developing will be made, we consider appropriate to start remembering the **§1** beginning of paper [1]:

"Let us take a system of co-ordinates in which the equations of Newtonian mechanics hold good. In order to render our presentation more precise and to distinguish this system of co-ordinates verbally from others which will be introduced hereafter, we call it the 'stationary system'.
If a material point is at rest relatively to this system of co-ordinates, its position can be defined relatively thereto by the employment of rigid standards of measurements and the methods of Euclidean geometry, and can be expressed in Cartesian co-ordinates."

In the following we will use Newtonian gravitational potential concept and Euclidean geometry, using polar coordinates for the central gravitational field considered.

Let be two material points **M** and **m** (one with a great mass **M**, and the other with a small mass **m<<M**). We can consider then **M** practically the **Centre of Mass** (**CM**) of the 2-point system (for example, **M** and **m** can model the **Earth** and an **electron**). In the corresponding **CM** inertial system, let be **r** the distance between **M** and **m**.

Let be

$$PE(r) = m_0(r)c^2 \qquad (4.1)$$

the **Potential Energy** (**PE**) of **m** measured by its **Rest Mass** $m_0$ (**c** is the vacuum light speed). We know that the gravitational **PE** increases when **r** increases. Its limit maximal value when **r** tends to infinite is then $m_{0m}c^2$, where $m_{0m}$ is the corresponding limit maximal value of the **Rest Mass** $m_0$. We have then

$$PE(r) = m_0(r)c^2 = m_{0m}c^2 - (GM/r)m_0(r) \qquad (4.2)$$

Here **G** is the Newtonian gravitational constant, and – **(GM/r)** is the **gravitational potential** owed to **M** with a supposed arbitrary value **0** at **r infinite**. **PE(r)** takes here the very definite maximal value $m_{0m}c^2$ at that place. The **gravitational potential** is the gravitational **Potential Energy** per unit of mass, being in this case the variable **Rest Mass** $m_0$ that measures the **Potential Energy**. With some simple algebraic handling we obtain

$$m_0(r) = m_{0m} / (1+GM/rc^2) \qquad (4.3)$$

We have then derived from **1905R** how the **Rest Mass** of a small body changes as a function of its position **r** in the central gravitational field of a great gravitational mass **M** body. The arbitrary additive constant characteristic of **Potential Energy** disappears in **1905R**, appearing a **zero Potential Energy** point at **r=0** that is not arbitrary, but a consequence of **Rest Mass** measuring **Potential Energy**.

If **M** and **m** are the **Earth** and an **electron**, $m_{0m}$ is the ordinary rest mass of a free electron (its maximal value at **r infinite**).

## 5. Determining atomic clock behavior with 1905R.

Since 1913 N. Bohr's **H** model [7], it is known that the frequency emitted by an atom is proportional to the **Rest Mass** of the **electron** involved in the change of state. In Einstein's **General Relativity** (**GR**), the **Rest Mass** is supposed a constant an intrinsic electron attribute, justifying the change in frequency (inverse of time) with the curvature of the space-time provoked by a mass-energy **M**. The things in **1905R** are very much simple. Taking as the reference the frequency of a clock at **r** infinite, we must multiply it by the **1905R** factor

$$1/[1+(GM/rc^2)] \qquad (5.1)$$

to obtain the frequency at **r**. The corresponding **GR** factor is known to be

$$\sqrt{[1-(2GM/rc^2)]} \qquad (5.2)$$

The change of frequency predicted by **1905R** is very close to the **GR** one in all the range of practical **r** values (**$GM/rc^2<<1$**), in real experiments like the Pound&Rebka [8] one or in the today Global Positioning System (GPS) [3] operation.

The **1905R** formula applies for all values of **r** from **0** to **infinite**, while the **GR** one doesn't apply for **$r<2GM/c^2$** where the factor takes imaginary values. It is not clear at a first glance which of the two approaches models Nature in a better way.

## 6. Conclusions.

Starting with a correct interpretation of the more than a century old paper [2] where the universal mass-energy relationship is derived, we had been able to explain atomic clock behavior in a gravitational field using only **1905 Relativity** (**1905R**), without the need to apply the later developed **Special Relativity** (**SR**) and **General Relativity** (**GR**).

It had been proved then that **GR** is not the unique valid alternative for the explanation of atomic clock behavior in a gravitational field, answering negatively the question exposed in the Introduction. Then, a new alternative road had been opened for the developing of a relativistic mechanics, related with the discovery that **Rest Mass** measures **Potential Energy**, a thing that seems to be not realized even by the one who does it: **1905 Albert Einstein**.

## 7. References.